\documentclass{article}
\usepackage{spconf,amsmath,graphicx}
\usepackage{multirow}
\usepackage[aboveskip=3pt, belowskip=2pt]{subcaption}
\usepackage{arydshln}
\usepackage{adjustbox, bm}

\usepackage{url,times}
\usepackage{color}
\usepackage{pifont}
\newcommand{\cmark}{\ding{51}}%
\newcommand{\xmark}{\ding{55}}%
\usepackage{amssymb}
\usepackage{booktabs}
\usepackage{graphics}
\usepackage{float,threeparttable}
\usepackage{commath}
\usepackage{mathtools, fixmath}
\usepackage{here}
\usepackage{cite}
\usepackage{array,etoolbox}
\preto\tabular{\setcounter{magicrownumbers}{0}}
\newcounter{magicrownumbers}

\usepackage{stackengine}
\stackMath

\usepackage{tabularx}
\title{Directional ASR: A new paradigm for E2E multi-speaker speech recognition with source localization}

\name{Aswin Shanmugam Subramanian$^\ast$, Chao Weng$^\dag$, Shinji Watanabe$^\ast$, Meng Yu$^\ddag$,}
\secondlinename{Yong Xu$^\ddag$, Shi-Xiong Zhang$^\ddag$, Dong Yu$^\ddag$}
\address{$^\ast$Center for Language and Speech Processing,
            Johns Hopkins University, Baltimore, MD, USA\\
         $^\dag$Tencent AI Lab, Shenzhen, China \quad \quad  $^\ddag$Tencent AI Lab, Bellevue, WA, USA
}

\begin{document}
\ninept

\maketitle

\begin{abstract}
This paper proposes a new paradigm for handling far-field multi-speaker data in an end-to-end neural network manner, called \textit{directional} automatic speech recognition (D-ASR), which explicitly models source speaker locations.
In D-ASR, the azimuth angle of the sources with respect to the microphone array is defined as a latent variable. This angle controls the quality of separation, which in turn determines the ASR performance. All three functionalities of D-ASR: localization, separation, and recognition are connected as a single differentiable neural network and trained solely based on ASR error minimization objectives.  
The advantages of D-ASR over existing methods are threefold: (1) it provides explicit speaker locations, (2) it improves the explainability factor, and (3) it achieves better ASR performance as the process is more streamlined. In addition, D-ASR does not require explicit direction of arrival (DOA) supervision like existing data-driven localization models, which makes it more appropriate for realistic data. For the case of two source mixtures, D-ASR achieves an average DOA prediction error of less than three degrees. It also outperforms a strong far-field multi-speaker end-to-end system in both separation quality and ASR performance.
\end{abstract}
\begin{keywords}
source localization, source separation, end-to-end speech recognition
\end{keywords}

\section{Introduction}
\label{sec:intro}
\vspace*{-2mm}

Source localization methods to estimate the direction of the sound sources with respect to the microphone array is an important front-end for many downstream applications. For example, they are an indispensable part of robot audition systems \cite{hark, nakadai2001real} that facilitates interaction with humans. Source localization is used as a pivotal component in the robot audition pipeline to aid source separation and recognition. Recently far-field systems are being designed to process multi-talker conversations like meeting \cite{yoshioka2019advances} and smart speaker scenarios \cite{chime5is,haeb2019speech}. Incorporating source localization functionality can enrich such systems by monitoring the location of speakers and also potentially help improve the performance of the downstream automatic speech recognition (ASR) task. Assuming the direction of arrival (DOA) is known, the effectiveness of using features extracted from the ground-truth location for target speech extraction and recognition was shown in \cite{zhuo_loc, icassp-tencent}.

Similar to other front-end tasks, signal processing has been traditionally used to estimate the DOA \cite{music_schmidt, tops, 6616137}. Wideband subspace methods like test of
orthogonality of projected subspaces (TOPS) have shown promising improvements over narrowband subspace methods like multiple signal classification (MUSIC) but most existing signal processing approaches are not robust to reverberations \cite{chakrabarty2019multi}. Supervised deep learning methods have been successful in making the DOA estimation more robust \cite{8553182, chakrabarty2019multi, nelson_localization}.  
However, for real multi-source data like CHiME-5 \cite{chime5is}, it is very difficult to get the parallel ground-truth DOA. In such cases, it might be possible to use labels from a different view that are easier to annotate and use it to indirectly optimize the localization model parameters.

There is growing interest in optimizing the front-end speech processing systems with applications-oriented objectives. For example, speech enhancement and separation systems have been trained based on ASR error minimization \cite{seltzer2003microphone, narayanan2014joint, Ochiai2017MultichannelES, beamnet, settle2018end, aswin_waspaa2019, chang2019mimo, icassp-tencent}. The front-end in such systems was encompassed into the ASR framework to train them without parallel clean speech data by using the text transcription view as the labels. MIMO-Speech \cite{chang2019mimo} is a multichannel end-to-end (ME2E) neural network that defines source-specific time-frequency (T-F) masks as latent variables in the network which in turn are used to transcribe the individual sources. 
 
Although MIMO-Speech might implicitly learn localization information, it might not be consistent across frequencies because of narrowband approximation. We propose to further upgrade the explainability factor of the MIMO-Speech system by realizing an explicit source localization function. 
We call our novel technique \textit{directional ASR} (D-ASR), which is a new paradigm of joint separation and recognition explicitly driven by source localization. 
The masking network in MIMO-Speech is expected to directly predict an information rich T-F mask. It is hard to accurately estimate such a mask without the reference signals. In D-ASR, this masking network is replaced with a simpler component that discretizes the possible DOA azimuth angles and estimates the posterior of these angles for each source. This estimated angle can be in turn converted to steering vectors and T-F masks, thereby making the localization function tightly coupled with the other two functionalities of D-ASR. This streamlining makes this method more effective. The estimated angle is tied across the steering vector for all frequency bands and hence this method also has wideband characteristics. 

Although D-ASR is trained with only the ASR objective, its evaluation is performed across all three of its expected functionalities. This is possible as the localization and separation intermediate outputs are interpretable in our formulation. The L1 prediction error is used as the DOA metric with MUSIC and TOPS as baselines for this comparison. Source separation and ASR performance are compared with the MIMO-Speech model \cite{mimo_transformer} with objective signal quality and transcription based metrics, respectively.

\vspace*{-2mm}
\section{Directional ASR}
\label{sec:baseline}
\begin{figure}[tb]
\begin{centering}
\includegraphics[scale=0.18]{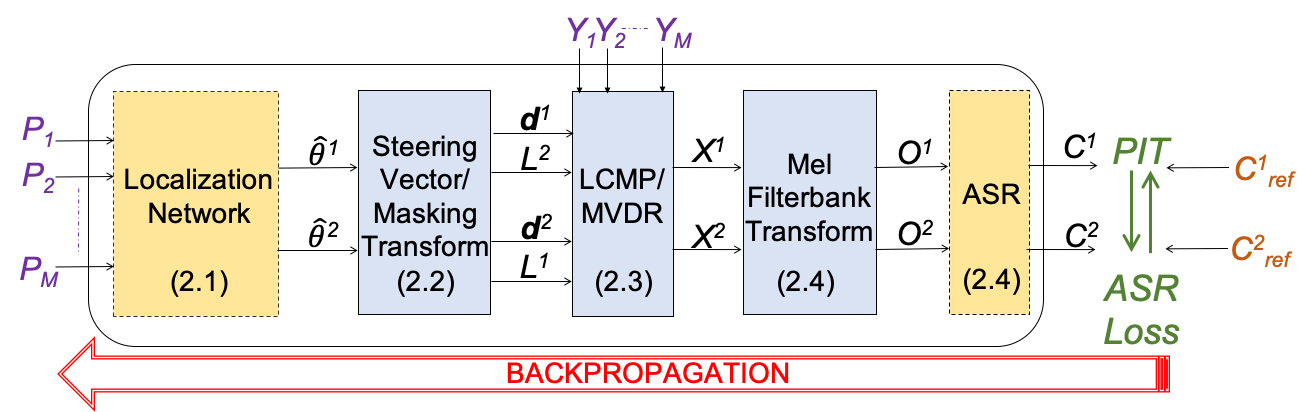}
\caption{Proposed D-ASR architecture for 2-speaker scenario. The DOA estimates $\hat{\theta}^{1}$ and $\hat{\theta}^{2}$ corresponding to the two sources are obtained as intermediate outputs. The yellow blocks have learnable parameters.}
\label{fig:EE}
\end{centering}
\vspace*{-2mm}
\end{figure}

\vspace*{-2mm}
A block diagram of the proposed D-ASR architecture is shown in Figure \ref{fig:EE}. The architecture has five blocks and a detailed formulation of these blocks is given in this section.
\vspace*{-2mm}
\subsection{Localization Subnetwork}
\vspace*{-2mm}
\label{sec:loc_net}
The first block is the localization subnetwork and its goal is to estimate the azimuth angle of the sources. This block in turn has three components: (1) CNN-based phase feature extraction, (2) phase feature masking and, (3) temporal averaging. 

Let $Y_1, Y_2, \cdots, Y_M$ be the $M$-channel input signal in the short-time Fourier transform (STFT) domain, with $Y_m \in \mathbb{C} ^{T\times F}$, where $T$ is the number of frames and $F$ is the number of frequency components. The input signal is reverberated, consisting of $N$ speech sources with no noise. We assume that $N$ is known. The phase spectrum of the multichannel input signal is represented as $\mathcal{P} \in [0, 2 \pi] ^{T\times M \times F}$. 
This raw phase $\mathcal{P}$ is passed through the first component of the localization network based on CNN given by $\text{LocNet-CNN}(\cdot)$ to extract phase feature $Z$ by pooling the channels as follows:
\vspace*{-1mm}
\begin{align}
     Z &= \text{LocNet-CNN} (\mathcal{P}) \in \mathbb{R} ^{T\times Q},
    \label{locnet_cnn}
\end{align}
where $Q$ is the feature dimension. Phase feature $Z$ will have DOA information about all the sources in the input signal. This is processed by the next component $\text{LocNet-Mask}(\cdot)$ which is a recurrent layer. This component is used to extract  source-specific binary masks as follows,
\vspace*{-1mm}
\begin{align}
     [W^{n}]_{n=1}^{N} &= \sigma(\text{LocNet-Mask} (Z)),
     \label{locnet_mask}
\end{align}
where $W^{n} \in [0, 1]^{T \times Q}$ is the feature mask for source $n$ and $\sigma(\cdot)$ is the sigmoid activation. This mask segments $Z$ into regions that correspond to each source.

In this work, we assume that the DOA does not change for the whole utterance. 
So in the next step we use the extracted phase mask from Eq.~\eqref{locnet_mask} to perform a weighted averaging of the phase feature from Eq.~\eqref{locnet_cnn} to get source-specific summary vectors. This summary vector should encode the DOA information specific to a source and that is why the mask is used as weights to summarize information only from the corresponding source regions.

\begin{align}
\vspace*{-1mm}
     \xi^{n} (q) &= \frac{\sum_{t=1}^{T} w^{n} (t, q) z(t, q)}{\sum_{t=1}^{T} 
     w^{n} (t, q)},
\end{align}
where $\xi^{n} (q)$ is the summary vector for source $n$ at dimension $q$, $w^{n}(t, q) \in [0, 1]$ and $z(t, q) \in \mathbb{R}$ are the extracted feature mask (for source $n$) and the phase feature, respectively, at time $t$ and feature dimension $q$. 

The summary vector, represented in vector form as $\bm{\xi}^{n} \in \mathbb{R} ^{Q}$ is passed through a learnable $\text{AffineLayer}(\cdot)$, which converts the summary vector from dimension $Q$ to the dimension $\lfloor 360/\gamma \rfloor$, where $\gamma$ is the angle resolution in degrees to discretize the DOA angle.
Based on this discretization, we can predict the DOA angle as a multi-class classifier with the softmax operation.
From this, we can get the source-specific posterior probability for the possible angle classes as follows,
\vspace*{-1mm}
     
\begin{align}
     \label{posterior}
     [\Pr (\theta^{n}=\alpha_{i}|\mathcal{P})]_{i=1}^{\lfloor 360/\gamma \rfloor} &= \text{Softmax}(\text{AffineLayer}(\bm{\xi}^{n})), \\
     \label{interpolation}
     \hat{\theta}^{n} &= \sum_{i=1}^{\lfloor 360/\gamma \rfloor} \Pr (\theta^{n}=\alpha_{i}|\mathcal{P}) \alpha_{i}, \\
     \alpha_{i} &=  \left( (\gamma * i) -\left((\gamma-1)/2\right) \right) \left(\pi/180\right),
     \label{eq:expected_doa}
\end{align}
where $\hat{\theta}^{n}$ is the DOA estimated for source $n$ by performing a weighted sum using the estimated posteriors from Eq.~\eqref{posterior} and $\alpha_{i}$ is the angle in radians corresponding to the class $i$. 
We define the composite function of Eqs.~\eqref{locnet_cnn}--\eqref{eq:expected_doa} with learnable parameter $\Lambda _{\text{loc}}$ as follows:
\vspace*{-1mm}
\begin{equation}
 [\hat{\theta}^{n}]_{n=1}^{N} = \text{Loc}(\mathcal{P}; \Lambda _{\text{loc}}).
 \label{eq:locnet}
\end{equation}
Note that all of the functions in this section are differentiable.

\subsection{Steering Vector \& Localization Mask}
\vspace*{-2mm}
In this step, the estimated DOAs from Eq.~\eqref{eq:locnet} is converted to a steering vector and also optionally to a time-frequency mask to be used for beamforming. 
First, the steering vector $\mathbf{d}^{n} (f) \in \mathbb{C} ^{M}$ for source $n$ and frequency $f$ is calculated from estimated DOA $\hat{\theta}^{n}$. 
In this work we have used uniform circular arrays (UCA) and the steering vector is calculated as follows,
\vspace*{-1mm}
\begin{align}
     \tau_{m}^{n} &= \frac{r}{c} \cos({\hat{\theta}^{n} - \psi_{m}}),\quad m=1:M \\
     \mathbf{d}^{n} (f) &= [e^{j2\pi f \tau_{1}^{n}}, e^{j2\pi f \tau_{2}^{n}}, ..., e^{j2\pi f \tau_{M}^{n}}],
     \label{eq:svector}
\end{align}
where $\tau_{m}^{n}$ is the signed time delay between the $m$-th microphone and the center for source $n$, $\psi_{i}$ is the angular location of microphone $m$, $r$ is the radius of the UCA and $c$ is the speed of sound (343 m/s). 

The estimated steering vectors can be sufficient to separate the signals.
However, depending on the beamformer used subsequently, we can also use a time-frequency mask $l^{n}(t,f) \in [0, 1)$ and it is estimated as, 

\vspace*{-1mm}
\begin{align}
a^{n}(t,f) &= |\mathbf{d}^{n} (f)^\textrm{H} \mathbf{y} (t,f)|^{2}, \\
\label{eq:doa-mask-pre}
[\nu^{n}(t, f)]_{n=1}^{N} &= \text{Softmax}([a^{n}(t ,f)]_{n=1}^{N}), \\
l^{n}(t, f) &= \frac{1}{(1-\kappa)} * \text{ReLU}(\nu^{n}(t ,f)-\kappa),
\label{eq:doa-mask}
\end{align}

where $\kappa \in [0,1)$ is a sparsity constant, $^\textrm{H}$ is conjugate transpose, $\mathbf{y} (t,f) \in \mathbb{C} ^{M}$ is the multichannel input at time $t$ and frequency $f$. An initial estimate of the mask is extracted by passing the directional power spectrum $a^{n}(t ,f)$ through a softmax function over the source dimension using Eq.~\eqref{eq:doa-mask-pre}. The mask is further refined  using Eq.~\eqref{eq:doa-mask} to create sparsity and we define this output as the \textit{localization mask}.
Again, all of the operations in this section are differentiable.

\subsection{Differentiable Beamformer}
\vspace*{-2mm}
As we have both steering vectors from Eq.~\eqref{eq:svector} and masks from Eq.~\eqref{eq:doa-mask}, there are many possible beamformers that could be used. We experimented with three options. First, linearly constrained minimum power (\textbf{LCMP}) beamformer \cite{gannot_consolidated} given by $\mathbf{b}_{\mathsf{LCMP}}^n (f) \in \mathbb{C} ^M $ for source $n$ and frequency $f$, is estimated by solving the following equation which requires only the steering vectors,
\vspace*{-1mm}
\begin{align}
\label{estimate_lcmp}
\mathbf{b}_{\mathsf{LCMP}}^n (f) &= \mathbf{\Phi}_{\text{y}}(f)^{-1}\mathbf{G}(f)[\mathbf{G}(f)^\textrm{H}\mathbf{\Phi}_{\text{y}}(f)^{-1}\mathbf{G}(f)]^{-1} \bm{\mu}^n, \\
\mathbf{\Phi}_{\text{y}} (f) &= \frac{1}{T}\sum\limits_{t=1}^T \mathbf{y}(t, f)\mathbf{y}(t, f)^\textrm{H},
\end{align}
where $\mathbf{\Phi}_{\text{y}} (f) \in \mathbb{C}^{M \times M}$ is the input spatial covariance matrix (SCM) at frequency $f$.
$\mathbf{G}(f) \in \mathbb{C}^{M \times N}$ is the constraint matrix whose columns are the estimated steering vectors at frequency $f$, such that the $n$-th column is $\mathbf{d}^{n} (f)$.
$\bm{\mu}^n \in \{0, 1\} ^{N}$ is a one-hot vector with the $n$-th element as 1. Alternative to LCMP, we can use \textbf{MVDR} beamformer. The localization masks from Eq.~\eqref{eq:doa-mask} are used to compute the source specific SCM, $\mathbf{\Phi}^{n} (f)$  as follows:
\vspace*{-1mm}
\begin{align}
\mathbf{\Phi}^{n} (f) & = \frac{1}{\sum_{t=1}^T l^{n}(t,f)}\sum\limits_{t=1}^T l^{n}(t,f)\mathbf{y}(t, f)\mathbf{y}(t, f)^\textrm{H}, 
\label{psd}
\end{align}
The interference SCM, $\mathbf{\Phi}_{\mathsf{intf}}^{n} (f)$ for source $n$ is approximated as $\sum_{i\neq n}^N \mathbf{\Phi}^{i} (f)$ like \cite{chang2019mimo} (we experiment only with $N=2$, so no summation in that case). From the computed SCMs, the $M$-dimensional complex MVDR beamforming filter for source $n$ and frequency $f$, $\mathbf{b}_{\mathsf{MVDR}}^n (f) \in \mathbb{C} ^M$ is estimated as,
\vspace*{-1mm}
\begin{equation}
  \mathbf{b}_{\mathsf{MVDR}}^n (f) = \frac{[\mathbf{\Phi}_{\mathsf{intf}}^{n} (f)]^{-1} \mathbf{d}^{n} (f)}{ \mathbf{d}^{n} (f)^{\textrm{H}}[\mathbf{\Phi}_{\mathsf{intf}}^{n} (f)]^{-1}\mathbf{d}^{n} (f)}.
\label{estimate_mvdr}
\end{equation}
We can also use the MVDR formulation based on reference selection \cite{mvdr_souden}, in which case the steering vectors will not be used explicitly. We refer to this beamformer as \textbf{MVDR-REF} and it is estimated as,
\vspace*{-1mm}
\begin{equation}
\mathbf{b}_{\mathsf{MVDR-REF}}^n (f) = \frac{[\mathbf{\Phi}_{\mathsf{intf}}^{n} (f)]^{-1} \mathbf{
\Phi}^{\text{n}} (f)}{\text{Tr}([\mathbf{\Phi}_{\mathsf{intf}}^{n} (f)]^{-1}\mathbf{\Phi}^{\text{n}} (f))} \mathbf{u},
\label{estimate_mvdr_ref}
\end{equation}
where $\mathbf{u} \in \{0, 1\} ^{M}$ is a one-hot vector to choose a reference microphone and $\text{Tr}(\cdot)$ denotes the trace operation.

Once we obtain the beamforming filters from either Eq.~\eqref{estimate_lcmp}, Eq.~\eqref{estimate_mvdr} or Eq.~\eqref{estimate_mvdr_ref}, we can perform speech separation to obtain the $n$-th separated STFT signal, $x^{n} (t, f) \in \mathbb{C}$ as follows:
\vspace*{-1mm}
\begin{equation}
 x^{n} (t, f) = \mathbf{b}^n (f)^\textrm{H} \mathbf{y}(t, f),
\label{apply_bf}
\end{equation}
where $\mathbf{b}^n (f)$ can be either of LCMP, MVDR or MVDR-REF beamforming coefficients.
Again, all the operations in this section are also differentiable. One added advantage of our approach is that we can use a different beamformer during inference to what was used during training. Irrespective of the beamformer used while training, we found that using MVDR-REF while inference gives better separation and ASR performance.

\subsection{Feature Transformation \& ASR}
\vspace*{-2mm}
\label{sec:asr}
\begin{align}
    O^{n} & = \text{MVN}(\text{Log}(\text{MelFilterbank}(\abs{X^{n}})))\\
    C^{n} & = \text{ASR}(O^{n}; \Lambda_{\text{asr}}).
    \label{eq:asrnet}
\end{align}
The separated signal for source $n$ from Eq.~\eqref{apply_bf}, represented in matrix form as $X^{n} \in \mathbb{C}^{T \times F}$ is transformed to a feature suitable for speech recognition by performing log Mel filterbank transformation and utterance based mean-variance normalization (MVN). The extracted feature $O^{n}$ for source $n$ is passed to the speech recognition subnetwork $\text{ASR}(\cdot)$ with learnable parameter $\Lambda_{\text{asr}}$ to get $C^{n} = (c^{n}_{1}, c^{n}_{2}, \cdots)$, the token sequence corresponding to source $n$. 

As shown in Figure \ref{fig:EE}, all the components are connected in a computational graph and  trained solely based on the
ASR objective to learn both 
$\Lambda _{\text{loc}}$ in Eq.~\eqref{eq:locnet} and $\Lambda _{\text{asr}}$ in Eq.~\eqref{eq:asrnet} with the reference text transcriptions $[C_{ref}^{i}]_{i=1}^{N}$ as the target. The joint connectionist temporal classification (CTC)/attention loss \cite{kim2016joint_icassp2017} is used as the ASR optimization criteria. We use the permutation invariant training (PIT) scheme similar to MIMO-Speech \cite{chang2019mimo, mimo_transformer} to resolve the prediction-target token sequence assignment problem.  Optionally, a regularization cross entropy loss for the posteriors in  Eq.~\eqref{posterior} with a uniform probability distribution as the target can be added. This is to discourage a very sparse posterior distribution and encourage interpolation in Eq.~\eqref{interpolation}, especially when high values of $\gamma$ are used.

\section{Experiments}
\vspace*{-2mm}

\subsection{Data \& Setup}
\vspace*{-2mm}
We simulated 2-speaker mixture data using clean speech from the subset WSJ0 of the wall street journal (WSJ) corpus \cite{paul1992design}. For each utterance, we mixed another utterance from a different speaker within the same set, so the resulting simulated data is the same size as the original clean data with 12,776 (si\_tr\_s), 1,206 (dt\_05), and 651 (et\_05) utterances for the training, development, and test set respectively. The SMS-WSJ \cite{SmsWsj19} toolkit was used for creating the simulated data with maximum overlap. Image method \cite{allen1979image} was used to create the room impulse responses (RIR). Room configurations with the size (length-width-height) ranging from 5m-5m-2.6m to 11m-11m-3.4m were used. A uniform circular array (UCA) with a radius of 5 cm was used. The reverberation time (T60) was sampled uniformly between 0.15s
and 0.5s. The two speech sources were placed randomly at a radius of 1.5-3m around the microphone-array center. 

Three CNN layers with rectified linear unit (ReLU) activation followed by a feedforward layer were used as $\text{LocNet-CNN}(\cdot)$ defined in Eq.~\eqref{locnet_cnn}. $Q$ was fixed as ``$2 \times \lfloor 360/\gamma \rfloor$".  One output gate projected
bidirectional long short-term memory (BLSTMP) layer with $Q$ cells was used as $\text{LocNet-Mask}(\cdot)$ defined in Eq.~\eqref{locnet_mask}. $\kappa$ in Eq.~\eqref{eq:doa-mask} was fixed as ``0.5". The encoder-decoder ASR network was based on the Transformer architecture \cite{espnet_transformer} and it is initialized with a pretrained model that used single speaker training utterances from both WSJ0 and WSJ1. Attention/CTC joint ASR decoding was performed with score combination with a word-level recurrent language model from \cite{wordlm} trained on the text data from WSJ. Our implementation was based on ESPnet \cite{espnet}.

 The input signal was preprocessed with weighted prediction error (WPE) \cite{wpe, Drude2018NaraWPE} based dereverberation with a filter order of ``10" and prediction delay ``3", only during inference time . The second channel was fixed as the reference microphone for MVDR-REF in Eq.~\eqref{estimate_mvdr_ref}. The recently proposed MIMO-Speech with BLSTM front-end and Transformer back-end \cite{mimo_transformer} was used as the 2-source ASR baseline. We used the same ASR subnetwork architecture as D-ASR in MIMO-Speech for a fair comparison. Signal processing based DOA estimation was performed with ``Pyroomacoustics" toolkit \cite{pra}.

\vspace*{-1mm}
\subsection{Multi-Source Localization Performance}
\vspace*{-2mm}
The estimated azimuth angles (DOA) from our proposed D-ASR method are extracted as intermediate outputs from Eq.~\eqref{eq:locnet}. We use two popular subspace-based signal processing methods MUSIC and TOPS as baselines. For both, the spatial response is computed and the top two peaks are detected to estimate the DOA. The average absolute cyclic angle difference between the predicted angle and the ground-truth angle in degrees is used as the metric. The permutation of the prediction with the reference that gives the minimum error is chosen.

The results with and without WPE preprocessing  are given in Table \ref{tab:doa_error}. Results of the D-ASR method with all three possible beamformers while training and different angle resolutions ($\gamma$) are shown. All configurations of D-ASR outperforms the baselines. Specifically, LCMP version is significantly better. In the LCMP version of D-ASR, having a higher $
\gamma$ of 10 works best and it is also robust without WPE. The results of LCMP D-ASR with $\gamma=5$ is shown with and without the additional regularization loss  (cross-entropy with uniform distribution as target). Although this regularization doesn't change the performance much, it helps in faster convergence during training.

\begin{figure}[tb]
\vspace*{-4mm}
\begin{centering}
\includegraphics[scale=0.5]{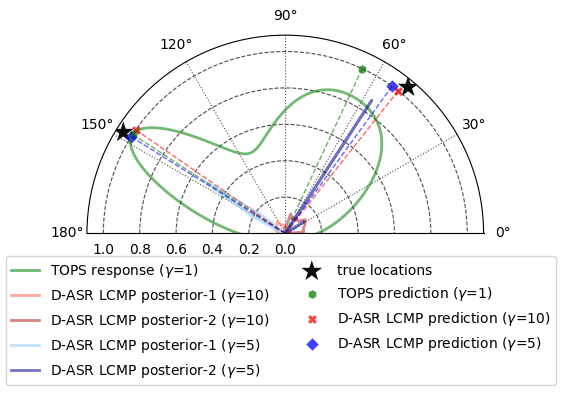}
\caption{Localization output given by D-ASR \& TOPS for a test example. Only the first two quadrants are shown. D-ASR LCMP with $\gamma=5$ here is without uniform regularization}
\label{fig:DOA_response}
\end{centering}
\vspace*{-2mm}
\end{figure}

If the posteriors in Eq.~\eqref{posterior} were 1-hot vectors, there would have been a discretization error of $5^{\circ}$ (for $\gamma$=10) but our prediction error is less than this. This shows that the network uses the posteriors as interpolation weights and hence it is not bounded by the discretization error. Figure \ref{fig:DOA_response} shows the localization output for a test example (note that only half of the possible angles are shown as in this case all the predictions were in the top semi-circle). TOPS predicts one of the sources which is at $148^{\circ}$ with good precision like the two D-ASR LCMP systems shown but has about a $15^{\circ}$ error in predicting the other source at $50^{\circ}$, which the D-ASR systems can estimate again with good precision. The posteriors of D-ASR ($\gamma=5$) trained without the regularization loss predicts a very sparse distribution while for the one with regularization loss and $\gamma=10$, the distribution is very dense but the prediction with the weighted sum is a bit more accurate.

\begin{table}[tb]
  \centering
  \caption{DOA Prediction Error on our simulated 2-source mixture comparing our proposed D-ASR method with subspace methods}
   \begin{adjustbox}{width=\columnwidth,center}
    \begin{tabular}{c|c|c|c|c|c|c|c}
    \toprule
    \multirow{3}[2]{*}{\textit{\textbf{Method}}} & \multirow{2}[1]{*}{\textit{\textbf{Training}}} &
    \multirow{3}[2]{*}{\textit{\textbf{$\gamma$}}} &
    \textit{\textbf{Uniform}} &\multicolumn{4}{c}{\textit{\textbf{Average L1 Error (degree)}}} \\
&   \multirow{2}[1]{*}{\textit{\textbf{Beamformer}}}     &       & \textit{\textbf{Regulariza-}} & \multicolumn{2}{c|}{\textit{\textbf{No WPE}}} & \multicolumn{2}{c}{\textit{\textbf{WPE}}} \\
&        &       & \textit{\textbf{tion}} & \textit{\textbf{Dev}} & \textit{\textbf{Test}} & \textit{\textbf{Dev}} & \textit{\textbf{Test}} \\
    \midrule
    MUSIC &  - & 1 & - & 22.5      &  24.0 & 15.6 & 14.4 \\
    MUSIC &  - & 10 &  - &  23.2    &  21.7 & 13.5 & 12.6 \\
    TOPS  & - & 1 & - &  \textbf{20.0}      & \textbf{18.7}    & \textbf{11.7}  & \textbf{10.1} \\
    TOPS  & - & 10 & - & 20.4         &   19.4  &  13.3 & 10.6  \\
    \midrule
    D-ASR   & LCMP & 1 & \cmark & 20.3 & 20.8 &  8.3    & 7.0    \\
    D-ASR   & LCMP & 5 & \cmark & 3.9 & 3.6 &  3.5    & 3.0    \\
    D-ASR   & LCMP & 5 & \xmark & 8.6 & 7.7 &  3.7    & 3.5    \\

    D-ASR   & LCMP & 10  & \cmark & \textbf{3.0} & \textbf{3.0} &    \textbf{2.5}   & \textbf{2.5}   \\
    D-ASR   & MVDR & 5 & \cmark & 11.5 & 10.4 &  7.4    &  7.1  \\
    D-ASR   & MVDR & 10 & \cmark &20.8 & 19.9 &   7.4    &  6.9  \\
    D-ASR   & MVDR-REF & 5  & \cmark& 14.2 & 14.5 &  10.2    &  10.9  \\
    \bottomrule
    \end{tabular}
     \end{adjustbox}
  \label{tab:doa_error}
\end{table}% 

\vspace*{-1mm}
\subsection{ASR \& Source Separation Performance}
\vspace*{-2mm}
Word error rate (WER) is used as the metric for ASR. The signal-to-distortion ratio (SDR) and perceptual evaluation of speech quality (PESQ) scores computed with the dry signal as the reference are used as the metric for source separation. The results are given in Table \ref{tab:wer}. The LCMP based D-ASR with $\gamma=10$ is used for this evaluation as that performs the best in terms of DOA error. We observed the steering vector based LCMP beamformer to be good at training the D-ASR network as that ensures tighter coupling with the localization subnetwork but once the network parameters are learned it was better to replace the beamformer to be mask-based for inference. So during inference MVDR-REF beamformer was used as mentioned in Section \ref{sec:asr}. 
\begin{table}[tb]
  \centering
  \caption{ASR \& Speech Separation Performance comparing our proposed D-ASR method with MIMO-Speech. For WER, lower the better and for SDR \& PESQ, higher the better. }
  \begin{adjustbox}{width=\columnwidth,center}
    \begin{tabular}{c|c|c|c|c|c|c}
    \toprule
     & \multicolumn{2}{c|}{\textit{\textbf{WER (\%)}}} & \multicolumn{2}{c|}{\textit{\textbf{SDR (dB)}}} & \multicolumn{2}{c}{\textit{\textbf{PESQ}}}  \\
& \textit{\textbf{Dev}} & \textit{\textbf{Test}} &  \textit{\textbf{Dev}} & \textit{\textbf{Test}} &  \textit{\textbf{Dev}} & \textit{\textbf{Test}}
\\
    \midrule
    Clean      & 2.1   & 1.6   &   -    &  -  &- &-\\
    Input mixture (ch-1) & 105.0 & 107.1 & -0.2 & -0.2 & 1.9 & 1.9 \\
    \midrule
    \textit{Oracle} binary Mask (IBM)    & 3.6 & 3.0   & 15.7  & 15.5 & 2.9 & 2.9 \\

    \textit{Oracle} DOA Mask (ILM)   & 4.3   & 3.7   &    15.3   &    15.3 & 2.9 & 2.9   \\
    \midrule
    MIMO-Speech  &  6.6  &  5.1 &  11.2     & 11.7  & 2.6 & 2.7 \\
    D-ASR   &  \textbf{5.1}     & \textbf{4.1}      &   \textbf{15.2}    &   \textbf{15.2}& \textbf{2.9} & \textbf{2.9}\textbf{}\\
\bottomrule  \end{tabular}
\end{adjustbox}%
  \label{tab:wer}%
\vspace*{-6mm}
\end{table}
The results of the clean single-speaker data with the pretrained ASR model is given in the first row to show a bound for the 2-mix ASR performance. The ASR results of the simulated mixture using the single-speaker model is also shown (WER more than 100\% because of too many insertion errors).  We perform oracle experiments by directly giving the reference masks to the beamformer. We define an ideal localization mask (ILM) which is basically Eq.~\eqref{eq:doa-mask} calculated by using the ground-truth DOA. The ILM is compared with the ideal binary mask (IBM) that is calculated with the reference clean signal magnitude. The ILM system is slightly worse than IBM but the results show it is a good approximation for IBM, which justifies our approach of obtaining masks from the DOA.

\begin{figure}[ht]
\vspace*{-2mm}
\begin{subfigure}{.5\columnwidth}
\centering
\includegraphics[width=4cm,height=2.5cm]{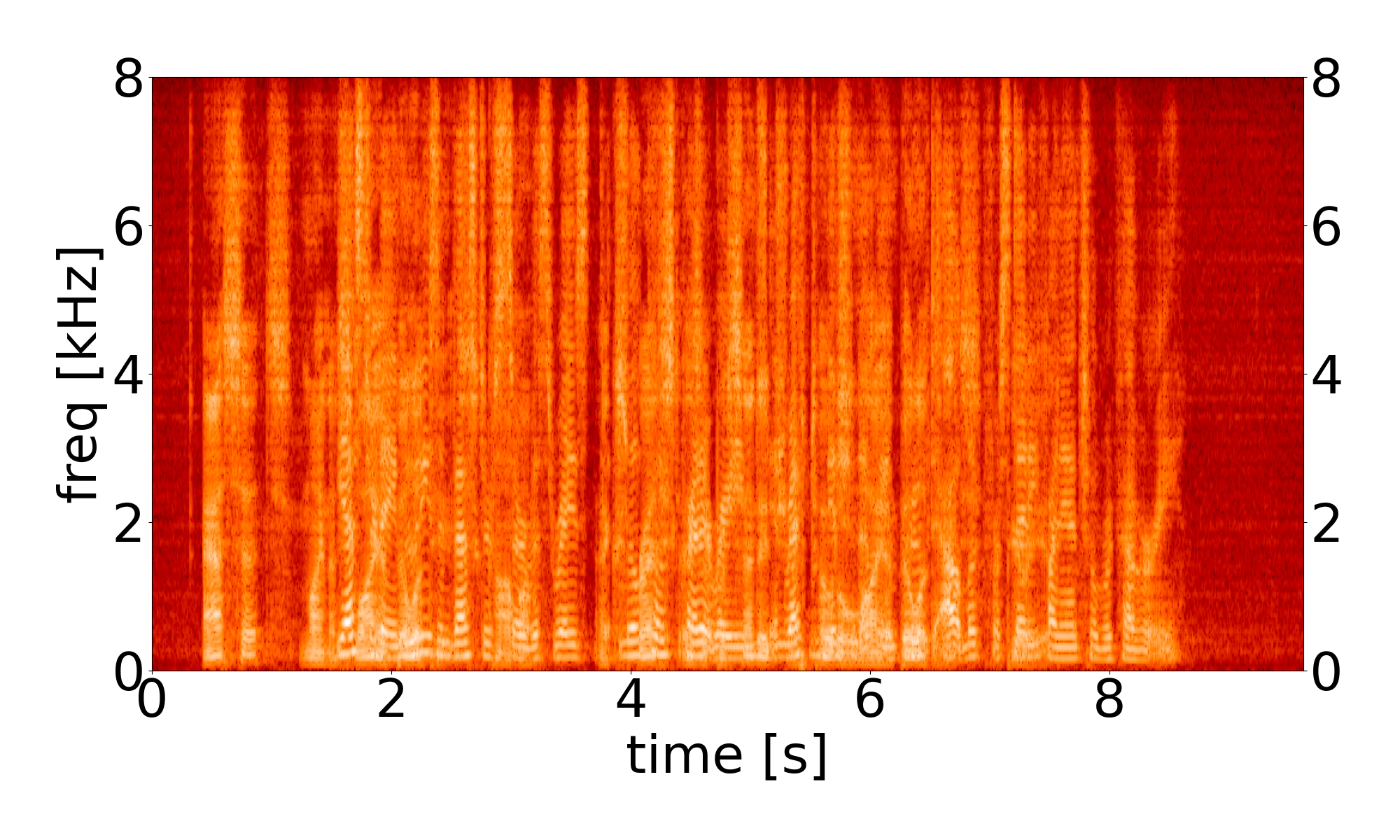}
\vspace*{-2mm}
\caption{Input overlapping speech}
\label{fig:sub1}
\end{subfigure}%
\begin{subfigure}{.5\columnwidth}
\centering
\includegraphics[width=4cm,height=2.5cm]{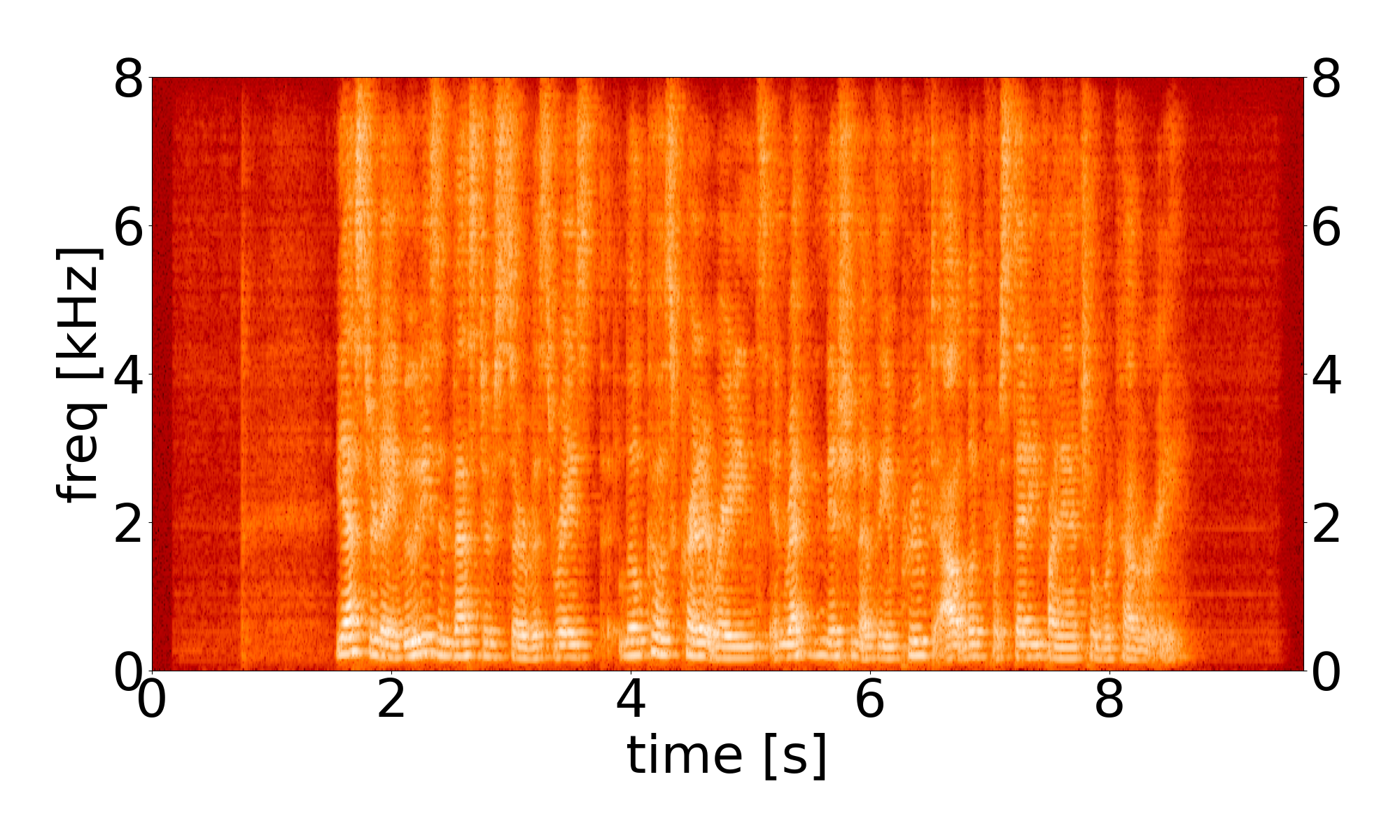}
\vspace*{-2mm}
\caption{Reference source-1}
\label{fig:sub2}
\end{subfigure}
\vspace*{-2mm}
\\[1ex]
\vspace*{-1mm}
\begin{subfigure}{.5\columnwidth}
\centering
\includegraphics[width=4cm,height=2.5cm]{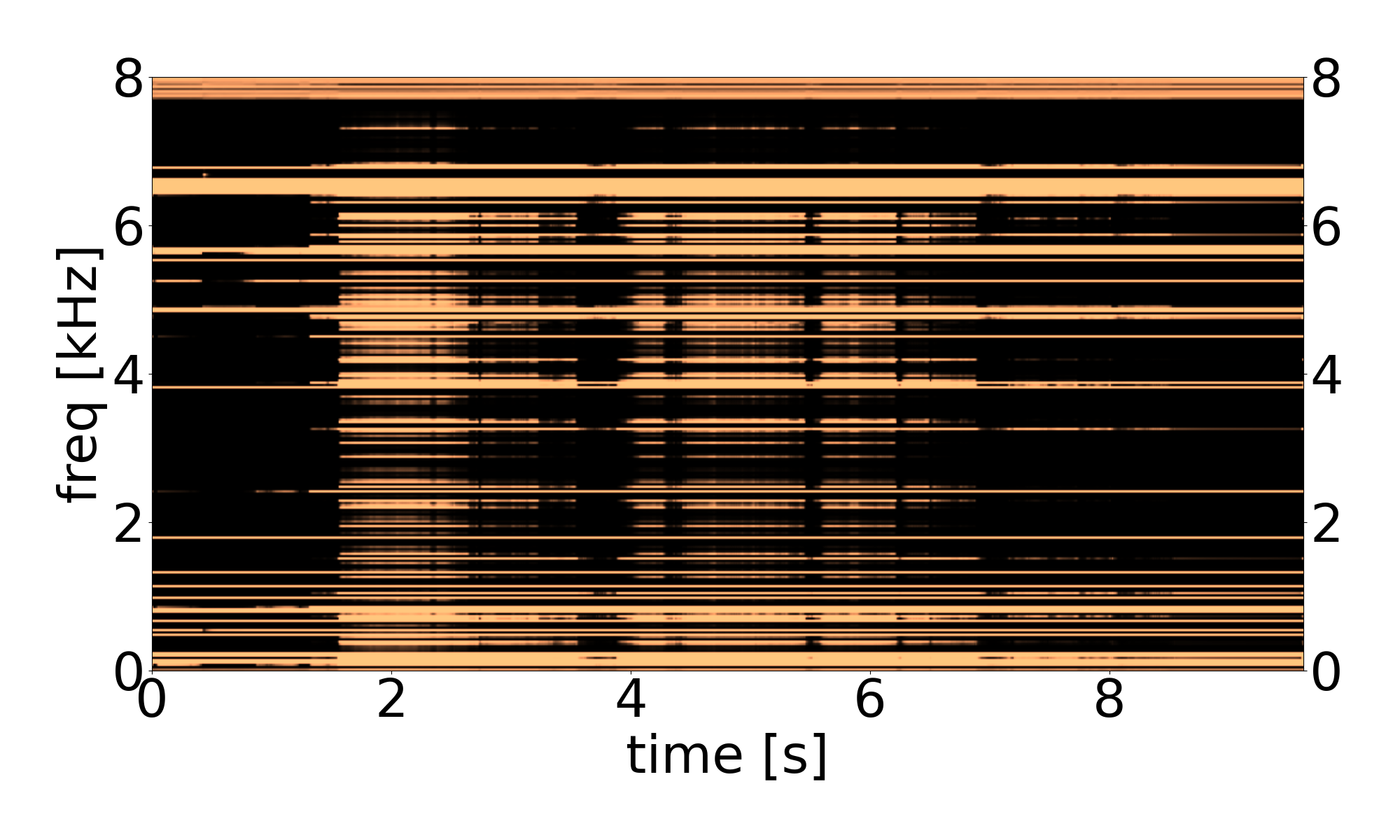}
\vspace*{-2mm}
\caption{MIMO-Speech Source-1 mask}
\label{fig:sub3}
\end{subfigure}
\begin{subfigure}{.5\columnwidth}
\centering
\includegraphics[width=4cm,height=2.5cm]{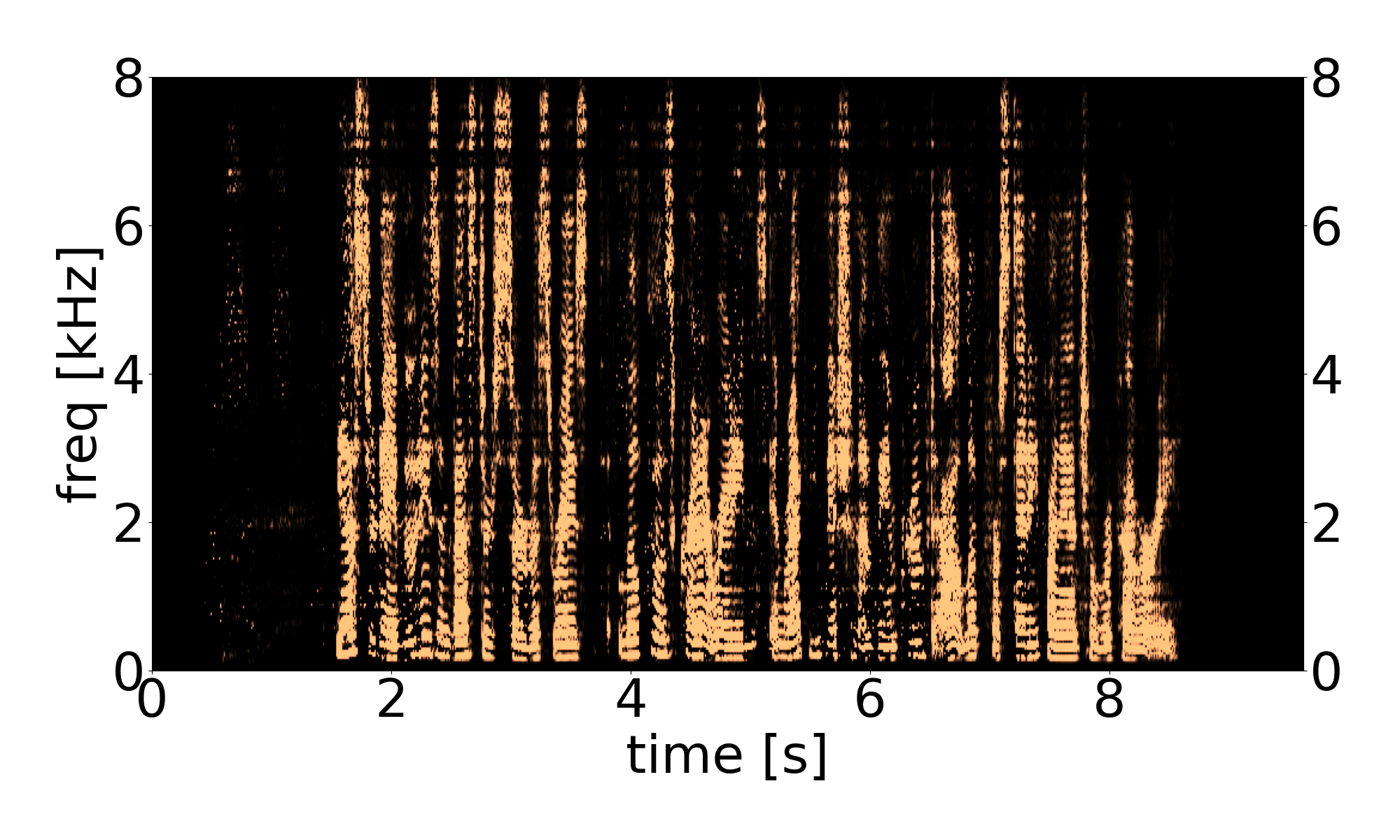}
\vspace*{-2mm}
\caption{D-ASR Source-1 Mask}
\label{fig:sub4}
\end{subfigure}

\caption{Intermediate T-F mask corresponding to one of the sources, estimated by D-ASR and MIMO-Speech for a test 2-spkr mix utterance in (a). The D-ASR mask in (d) captures the corresponding reference in (b) far better than the MIMO-Speech mask in (c).}
\label{fig:mask}
\vspace*{-1mm}
\end{figure}

Our proposed D-ASR method outperforms MIMO-Speech with the added advantage of also predicting DOA information. The SDR scores of D-ASR is almost as good as the oracle IBM and it is far superior to MIMO-Speech. An example of the estimated masks comparing D-ASR with MIMO-Speech is shown in Figure \ref{fig:mask}. In MIMO-Speech as the spectral masks are estimated directly, it is hard for the network to learn this without reference signals. Our approach can provide better masks because we simplify the front-end to just predict angles and then approximate the mask from it. Some audio examples for demonstration are given in \url{https://sas91.github.io/DASR.html}

\vspace*{-2mm}
\section{CONCLUSION}
\vspace*{-2mm}
\label{sec:conclusion}
This paper proposes a novel paradigm to drive far-field speech recognition through source localization. It also serves as a method to learn multi-source DOA from only the corresponding text transcriptions. D-ASR not only makes the existing approaches like MIMO-Speech more interpretable but also performs better in terms of both ASR and speech separation.  
\vfill\pagebreak
\label{sec:refs}
\bibliographystyle{IEEEbib}
\bibliography{refs}

\end{document}